\documentclass[aps,aps,twocolumn,superscriptaddress,groupedaddress,amsmath,amssymb]{revtex4}
\usepackage{graphicx}
\usepackage{color}
\usepackage{epsfig}
\usepackage{dcolumn}
\usepackage{bm}

\begin{document}

\title{Statistical theory of correlations in random packings of hard
  particles}

\author{Yuliang Jin} \affiliation {Levich Institute and Physics Department, City College of
  New York, New York, NY 10031, USA}
\author{James G. Puckett} \affiliation{Department of Mechanical Engineering and Materials Science, Yale University, New Haven, CT 06520, USA}
\author {Hern\'an A. Makse} \affiliation {Levich Institute and Physics Department, City College of
  New York, New York, NY 10031, USA}
\email{hmakse@lev.ccny.cuny.edu}


\begin{abstract}
A random packing of hard particles represents a fundamental model for granular matter.  Despite its importance, analytical modeling of random packings remains difficult due to the
existence of strong correlations which preclude the development of a simple theory. Here, we take inspiration from liquid theories for the $n$-particle angular correlation function to
develop a formalism of random packings of hard particles from the bottom-up. A progressive expansion into a shell of particles converges in the large layer limit under a Kirkwood-like
approximation of higher-order correlations. We apply the formalism to hard disks and predict the density of two-dimensional random close packing (RCP), $\phi_{\rm rcp} = 0.85\pm0.01$,
and random loose packing (RLP), $\phi_{\rm rlp} = 0.67\pm0.01$. Our theory also predicts a phase diagram and angular correlation functions that are in good agreement with experimental
and numerical data.


\end{abstract}

\maketitle

\clearpage

\section{Introduction}
In recent years, many important practical applications have been found for granular materials, which are commonly modeled by dense packings of hard spheres \cite{Aste2008Book}. Sphere
packing problems are equivalent to important problems in number theory and error-correcting coding \cite{Conway1999}, both of which are fundamental in computer science. Despite its
importance, analytical developments in granular matter have lagged behind in comparison with other fields of condensed matter, like liquid theory. In the case of random packings
\cite{Bernal1960A}, analytical results are still difficult to obtain. The theoretical difficulty arises due to {\it (i)} the absence of a first principle derivation of the statistical
ensemble of packings (such as Liouville's theorem in ordinary liquids) that would lead to a proper definition of randomness \cite{Torquato2000}, and {\it (ii)} the existence of
correlations between the particle positions determining the properties of random packings.



In previous theories, these correlations have been neglected or treated using simple approximations. For instance, Gotoh and Finney \cite{Gotoh1974} estimated the density of RCP based
only on correlations among the contact neighbors. Another example is the `granocentric' model \cite{Clusel2009}, which considers the correlations between the central particle and
nearest Voronoi neighbors. Beyond local correlations, the statistical treatment of Song {\it et al.} takes a mean-field approximation of the long-range correlations
\cite{Song2008,Jin2010B}. Other mean-field approaches are developed based on liquid theories \cite{Berryman1983} and replica theory (RT) of the glass transition \cite{Parisi2010}.
However, in low dimensions the effects of fluctuations are strong and mean-field approximations are insufficient. For example, in 2d, the coarse-grained approximation used in
Ref.~\cite{Song2008} works poorly~\cite{Meyer2010}, and therefore more sophisticated treatments of correlations become necessary.




%
In this paper, we aim to establish a framework for random packings that addresses the two problems stated above: {\it (i)} we define an ensemble of equiprobable graphs that satisfy the
jamming conditions to represent the statistics of all possible contact networks, and {\it (ii)} we take into account pair and higher-order
particle correlations that are important to describe low-dimensional systems. Inspired by the more advanced liquid theories, our formulation is analogous to the Yvon-Born-Green (YBG)
hierarchy \cite{Hansen1986} augmented to consider the contact network and local and global jamming conditions for packings.
We develop a systematic layer-expansion within a Kirkwood-like superposition approximation \cite{Hansen1986} to provide a phase diagram and predictions of the volume fractions of
jammed packings. The theoretical predictions on volume fractions and pair distribution functions agree well with experiments and computer simulations on two-dimensional frictional
packings. We also discuss the relation between the present approach and glass theory frameworks in search of unifications of random packings and glasses
\cite{Parisi2010,Krzakala2007,Mari2008,Mezard2011}.

The present approach builds up on the Edwards mean-field theory of packings developed by Song et al. \cite{Song2008}, by incorporating correlations between the particle positions.
Previous theory \cite{Song2008}  utilizes a mean-field assumption of uniformity of the particle density in the bulk as well as the particles in contact. The  present theory is a
bottom-up approach to take into account particle-particle correlations which were neglected in \cite{Song2008} in a systematic way. In the thermodynamic limit of infinite number of
particles in the bulk and contacts, the theory recovers the results of Song et al. \cite{Song2008}, namely the exponential form of the distribution of the excluded Voronoi volume which
is the basic result to predict the volume fraction of the packing.

The paper is organized as follows: in Sec.~\ref{sec:general} we develop a general theoretical formalism. The formalism is applied to 2d packings (Sec.~\ref{sec:2d}) which provides a
phase diagram (Sec.~\ref{sec:phase_diagram}). The theoretical predictions are tested with experiments and computer simulations in Sec.~\ref{sec:test}. At the end, we conclude our paper
with discussions (Sec.~\ref{sec:conclusion}).




\section{General formalism}
\label{sec:general}
\begin{figure}
\centerline{\hbox{\includegraphics[width=2.7in]{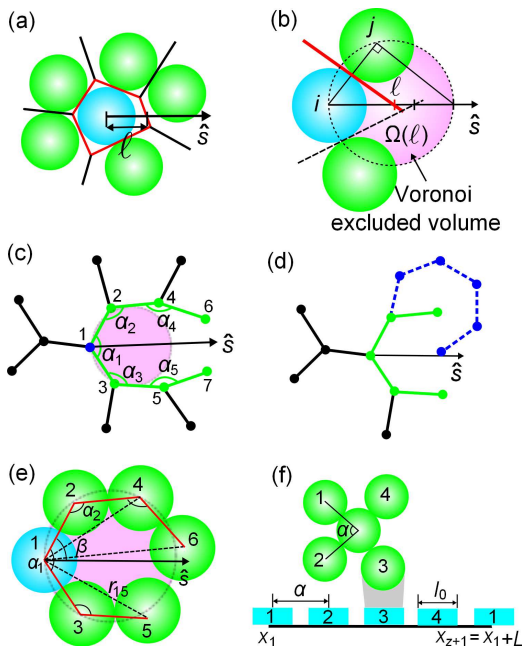}}}
 \caption{(Color online) {\bf Illustration of the theoretical formalism}. (a) A 2d illustration of the Voronoi volume
(bounded by red lines) and the Voronoi radius $\ell$ in direction $\hat{s}$. (b) The Voronoi excluded volume $\Omega(\ell)$ (pink area). $\ell$ is determined by particle $j$ because it
minimizes $\frac{r_{ij}}{2\hat{s} \cdot \hat{r}_{ij}}$.
(c) An illustration of the contact network and the Voronoi particles. Each dot represents a particle. The exclusive angle $\alpha_j$ is the angle between any two contact neighbors. No
other contact particles are allowed to be inside this angle. (d) In 2d, the Voronoi radius $\ell$ is determined by the Voronoi particles on the two closest branches (green). Other
particles may contribute only in an exceptional case such as shown by the dashed blue line. (e) An illustration of the geometrical quantities used in the calculation of $P(\ell)$. (f)
Mapping monodisperse contact disks to 1d rods. The 2d exclusive angle $\alpha$ corresponds to the 1d gap. } \label{fig:system}
\end{figure}

Within the context of Edwards statistical ensemble of packings \cite{Edwards1989,Song2008,Jin2010B, mbe, E21, sdr}, the volume associated to each particle plays
the role of the Hamiltonian, since packings tend to minimize the occupied volume rather than energy.  The fundamental quantity to describe the packing ensemble is the Voronoi volume
surrounding each particle,
which is defined as the volume of the Voronoi cell whose interior consists of the points that are closer to a given particle than to any other.
The $d$-dimensional Voronoi volume $W_i$ of particle $i$ is an angular average of a function of the ``Voronoi radius" $\ell$ \cite{Song2008} (see Fig.~\ref{fig:system}a):
\begin{equation}
W_i = \oint  \int _0^{\ell}  r^{d-1}d\hat{s}dr = \frac{S_d}{d}\langle \ell^d \rangle_s, \label{eq:wi}
\end{equation}
where $S_d$ is the $d$-dimensional solid angle. By the definition of Voronoi cell, $\ell$ is the minimum of the projection of the distance $\vec{r}_{ij}$ (from particle $i$ to any
other particle $j$) along the direction $\hat{s}$, $\ell \equiv \min_{\hat{s} \cdot
  \hat{r}_{ij}}\frac{r_{ij}}{2\hat{s} \cdot \hat{r}_{ij}}$, and
$\hat{s} \cdot \hat{r}_{ij}>0$ (Fig.~\ref{fig:system}b).

According to Eq.~(\ref{eq:wi}), the ensemble average of the Voronoi volume is:
\begin{equation}
\langle W \rangle_e =  \frac{1}{N} \sum_{i} \langle W_i \rangle_e  = \frac{S_d}{d} \langle \langle \langle \ell^d \rangle_s \rangle_i \rangle_e, \label{eq:wei}
\end{equation}
where $\langle \cdots \rangle_s$ is the average over direction $\hat{s}$, and $\langle \cdots \rangle_i$ is the average over particle $i$. In the random ensemble of homogeneous and
isotropic packings, each particle as well as each direction is equivalent. Thus $\ell$ is independent of particle $i$ and direction $\hat{s}$:
\begin{equation}
\langle W \rangle_e = \frac{S_d}{d} \langle \ell^d \rangle_e. \label{eq:we}
\end{equation}
Equation~(\ref{eq:we}) shows that it is enough to consider the distributions of particle positions along any arbitrary direction around any arbitrary particle, and the result is
representative for the global properties of the entire packing. This feature of random packings significantly simplifies the problem. Furthermore, this ensemble average can be
calculated from distribution functions:
\begin{equation}
\langle W \rangle_e  =\frac{S_d}{d} \int_0^\infty \ell^d p(\ell) d\ell  = -\frac{S_d}{d} \int_0^\infty \ell^d dP(\ell), \label{eq:W}
\end{equation}
and the packing fraction $\phi$ is the ratio between the volume of spheres
\begin{equation}
\phi = \frac{V_d}{\langle W \rangle_e}. \label{eq:voronoi}
\end{equation}
Here $p(\ell)$ is the probability distribution function of $\ell$ in the ensemble, and $P(\ell)$ is the inverse cumulative distribution function: $p(\ell) = -dP(\ell)/d\ell$.
According to the definition, $P(\ell)$ is the probability that $\frac{r_{ij}}{2\hat{s} \cdot
  \hat{r}_{ij}} > \ell$ for all $j$-particles at a distance $r_{ij}$
from $i$. Geometrically, $P(\ell)$ corresponds to the probability that all particles are outside a ``Voronoi excluded volume", $\Omega(\ell)$, which is a sphere of radius $\ell$
(Fig.~\ref{fig:system}b). The Voronoi excluded volume is a generalization of the excluded volume due to hard-core interactions, dating back to Onsager's hard rods solution
\cite{Onsager1949}. The distribution function $P(\ell)$ is similar to the exclusion probability function in the scaled particle theory for liquids \cite{Reiss1959}, and is related to
the $n$-particle correlation functions $g_n$ of all orders \cite{Jin2010B}.

As shown in Fig.~\ref{fig:system}c, to determine $P(\ell)$ we need to consider Voronoi particles which are the only ones with possible contributions to the Voronoi radius $\ell$.
This means that in the condition $\frac{r_{ij}}{2\hat{s} \cdot
  \hat{r}_{ij}} > \ell$ for $P(\ell)$, we only need to consider
particle $j$ labelled as a Voronoi particle.  In 2d, the Voronoi particles are located on the two closest branches to the direction $\hat{s}$, but
in higher dimensions more branches should be considered. The positions of the Voronoi particles are described by the $n$-particle angular correlation function $G_n(\alpha_{1},
\alpha_{2}, \ldots \alpha_{n})$ of exclusive angles.


However, to calculate $G_n$ one needs to define a proper ensemble first. Here we use the principle of entropy maximization which corresponds to a statistical treatment of an ensemble
of all jammed states, each of which has an equal probability \cite{Gotoh1974,Edwards1989,Song2008,Jin2010B}. This ensemble can be represented by a set of contact networks satisfying
the jamming condition, while for a given contact network, particle positions are allowed to fluctuate without destroying the contacts.
Our approach defines a random packing as the typical state in a flat average over the ensemble of all possible graphs of contact network configurations constraint to a given average
coordination number \cite{Song2008}.  Mechanical force and torque balance is assured by the isostatic condition imposed on the coordination number \cite{Alexander1998}.


The network representation is a unique feature of packings compared to unjammed liquid systems.
For contacting neighbors, we only need to know the distribution of the surface angles: the original $d$-dimensional problem is mapped onto a $(d-1)$-dimensional space. The theory is
mathematically treatable in two limits: {\it (i)} In 2d, the one-dimensional surface space can be analyzed analytically; {\it (ii)} In large dimensions, the contacting neighbors on the
surface can be approximated to the simple ideal gas \cite{Jin2010B}. Below we apply the general formalism to study 2d random packings, where correlations are more profound.

\section{Application of the theory in 2d}
\label{sec:2d} In this section, we apply the general formalism in 2d, and provide quantitative predictions which can be tested by experiments and computer simulations. The approach may
be generalized to higher dimensions, although the calculations might become much more complicated.

\subsection{Calculation of $P(\ell)$} \label{sec:pr}

By definition, $P(\ell)$ of the central particle $i=1$ is the probability that the Voronoi excluded volume $\Omega(\ell)$ is empty of particles, or equivalently $\frac{r_{1j}}{2\hat{s}
\cdot
  \hat{r}_{1j}} > \ell$ for any other particle $j$.
It is sufficient to only consider ``Voronoi particles" on the two closest branches except for the case shown in Fig.~\ref{fig:system}d. This exception disappears in the infinite
expansion order limit $n\rightarrow \infty$.
The condition that all Voronoi particles are outside $\Omega(\ell)$ requires that $\frac{r_{1j}}{2\cos \beta_j} > \ell$, where $r_{1i}$ is the distance between the central particle
$i=1$ and the Voronoi particle $j$, and $\cos \beta_j = \hat{s} \cdot \hat{r}_{1j}$. We can write $P(\ell)$ as:
\begin{equation}
\begin{split}
P(\ell)  = & \lim_{n' \rightarrow \infty} \int \cdots \int p(\vec{r}_{12}, \vec{r}_{13}, \ldots \vec{r}_{1n'} ) \\
&\times \prod_{j=2}^{n'} \Theta\left(\frac{r_{1j}}{2\hat{s} \cdot \hat{r}_{1j}} > \ell\right) d\vec{r}_{12} \cdots d\vec{r}_{1n'}, \label{eq:prr}
\end{split}
\end{equation}
where $n'$ is the total number of Voronoi particles considered, and $p(\vec{r}_{12}, \vec{r}_{13}, \ldots \vec{r}_{1n'} )$ is the distribution function of the positions of Voronoi
particles. The constraints $\Theta \left( \frac{r_{1j}}{2\hat{s} \cdot
  \hat{r}_{1j}} - \ell \right)$ impose the Voronoi exclusive
conditions. For a given contact network, the positions ($\vec{r}_{12}, \vec{r}_{13}, \ldots$) can be transformed to the exclusive angles ($\alpha_1, \alpha_2, \ldots$) and the angle
$\beta$ of the direction $\hat{s}$, see Appendix~\ref{sec:angles}.
Using this transformation, $P(\ell)$ becomes:
\begin{equation}
\begin{split}
P(\ell) = & \lim_{n \rightarrow \infty} \int \cdots \int p(\vec{r}_{12}, \vec{r}_{13}, \ldots \vec{r}_{1,n+2} )\\
&\times \prod_{j=2}^{n+2} \Theta\left(\frac{r_{1j}}{2\hat{s} \cdot
  \hat{r}_{1j}} > \ell\right)  \\
  &\times \frac{\partial(\vec{r}_{12},
  \vec{r}_{13}, \ldots \vec{r}_{1,n+2})}{\partial
  (\beta,\alpha_1,\ldots, \alpha_n)} d\beta d\alpha_1 \cdots
d\alpha_n, \label{eq:pra}
\end{split}
\end{equation}
where we let $n' = n +2$.
If the contact network is fixed, the degree of freedom of each particle is reduced from two to one.
Therefore the position variables $(\vec{r}_{12}, \vec{r}_{13}, \ldots \vec{r}_{1,n+2})$ and angular variables $(\beta,\alpha_1,\ldots, \alpha_n)$ have the same total $n+1$ degrees of
freedom.

Now the distribution of positions can be related to the distribution of angles:
\begin{equation}
\begin{split}
& p(\vec{r}_{12}, \vec{r}_{13}, \ldots \vec{r}_{1,n+2}) \frac{\partial(\vec{r}_{12}, \vec{r}_{13}, \ldots \vec{r}_{1,n+2})}{\partial (\beta,\alpha_1,\ldots, \alpha_n)} \\
&\sim G(\beta,\alpha_1,\ldots, \alpha_n) \\
& \sim \Theta(\alpha_{1} - \beta) G_n(\alpha_1,\ldots, \alpha_n). \label{eq:trans}
\end{split}
\end{equation}
The Heavyside function $\Theta(\alpha_{1} - \beta)$ means that the direction $\hat{s}$ is uniformly distributed and is bounded by the Voronoi particles ($\beta < \alpha_1$). Using
Eq.~(\ref{eq:trans}), we rewrite Eq.~(\ref{eq:pra}) with the $n$-particle angular correlation function $G_n$:




\begin{equation}
\begin{split}
P(\ell) = & \lim_{n \rightarrow \infty} \frac{z}{L} \int \cdots \int
\Theta(\alpha_{1} - \beta) G_n(\alpha_{1}, \ldots \alpha_{n}) \\
& \times \prod_{j=2}^{n+2}\Theta \left( \frac{r_{1j}}{2\hat{s} \cdot
  \hat{r}_{1j}} - \ell \right) d\beta d\alpha_{1} \cdots
d\alpha_{n}, \label{eq:PgIso}
\end{split}
\end{equation}
where $L = 2\pi$, $z$ is the average coordination number, and $z/L$ is a normalization factor determined from the condition that $P(1/2) = 1$ (we set the particle diameter to be one).
Equation~(\ref{eq:PgIso}) can be truncated at any value of $n$, and becomes exact in the limit $n \rightarrow \infty$. In this study, it is treated as an expansion of $n$ or number of
coordination layers ($n$ corresponds to twice of the number of layers).

Equation~(\ref{eq:PgIso}) is similar to the YBG hierarchy \cite{Hansen1986} in liquid theories in the sense that it relates one distribution function, $P(\ell)$, to another, $G_n$.
This similarity inspires us to bring the two approaches together to solve Eq.~(\ref{eq:PgIso}) within a closure approximation for $G_n$.  In liquid theory, the 3-point correlation
function $g_3$ is decomposed into the product of pair correlation functions $g_2$, by the use of Kirkwood's superposition approximation \cite{Kirkwood1935}.  This provides a closure of
the YBG hierarchy, which results in the non-linear integro-differential Born-Green equation \cite{Hansen1986}.  Here we use a similar Kirkwood-like approximation to decompose $G_n$
into the single-particle angular correlation function $G(\alpha)$:
\begin{equation} G_n(\alpha_{1}, \ldots \alpha_{n}) \approx
\prod_{j=1}^{n} G(\alpha_j). \label{eq:Gn} \end{equation}
This approximation neglects higher-order correlations between particles that do not share any common neighbors (Appendix~\ref{sec:kirkwood}).






\subsection{Calculation of the single-particle angular correlation function $G(\alpha)$ from a 1d model}

To find $G(\alpha)$, we map the contacting particles to a system of 1d rods with an effective potential.
As shown in Fig.~\ref{fig:system}f, the contact particles in 2d can be mapped to a set of $z$ interacting 1d hard rods at position $x_i$
of length $l_0 = \pi/3$
and system size $L = 2\pi =6 l_0$, with a periodic boundary condition.
The local jamming condition requires that each particle has at least $d+1$ contacting neighbors, and not all of these neighbors are in the same ``hemisphere''. In 2d, this means that
$z \geq 3$ and there is no exclusive angle $\alpha$ that could be greater than $\pi$. In the equivalent 1d model, the latter condition requires that no two nearest neighbors are
separated farther than $3l_0$. Thus, the jamming condition is equivalent to introducing an infinite square-well potential between two hard rods:
\begin{equation}
V(x)  = \begin{cases}
\infty, & \mbox{if } x/l_0<1 \mbox{ or } x/l_0>a\\
0,& \mbox{if } 1< x/l_0 < a,
\end{cases}
\label{eq:potential}
\end{equation}
with potential parameter $a = 3$. The total potential is a sum of the pairwise potentials,
\begin{equation}
\begin{split}
V(x_1, \cdots, x_z) =  &V(L-x_z) + V(x_z-x_{z-1}) + \cdots \\
&+ V(x_2-x_1).
\end{split}
\label{eq:V}
\end{equation}

To solve the 1d model, we first calculate the partition function $Q(L,z)$, which is
\begin{equation}
\begin{split}
Q(L,z)  = &\int \cdots \int \exp[- V(x_1, \cdots, x_z) \prod_{i=2}^{z} dx_i \\
= &\int_0^l \exp[-V(L-x_z)] dx_z \\
&\times \int_0^{x_z} \exp[-V(x_z-x_{z-1})] dx_{z-1} \cdots  \\
&\times \int_0^{x_3} \exp[-V(x_3-x_2)]\exp[- V(x_2)] dx_2,
\end{split}
\end{equation}
where we have used Eq.~(\ref{eq:V}), and set the temperature to be unit since it is irrelevant for our system. This integral is a $z$-fold convolution for the Laplace transform of the
function $\exp[-\beta
  V(x)]$ \cite{Salsburg1953}, which could be written as:
\begin{equation}
\begin{split}
Q(L,z) &= \frac{1}{2\pi i} \int_{\gamma - i \infty}^{\gamma + i
  \infty} e^{sL} q^z(s) ds, \\
   q(s) &= \int_0^{\infty}
\exp[-sx- V(x)]dx, \label{eq:partition2}
\end{split}
\end{equation}
where $\gamma$ is greater than the real parts of all the singularities of $q(s)$. If we plug the potential $V(x)$ (Eq.~(\ref{eq:potential})) in $q(s)$, we have (see
Appendix~\ref{sec:part} for details)
\begin{equation}
\begin{split}
Q(L, z)
 = &\sum_{k=0}^{\lfloor \frac{L/l_0-z}{2}\rfloor}(-1)^k \binom{z}{k}\frac{[L/l_0-z-2k]^{z-1}}{(z-1)!}\\
 &\times \Theta(L/l_0-z)\Theta(3z-L/l_0),
\end{split}
\label{eq:partition}
\end{equation}
where $\lfloor x \rfloor$ is the integer part of $x$.

\begin{figure}
\centerline{\hbox{ \includegraphics [width = 2.5in] {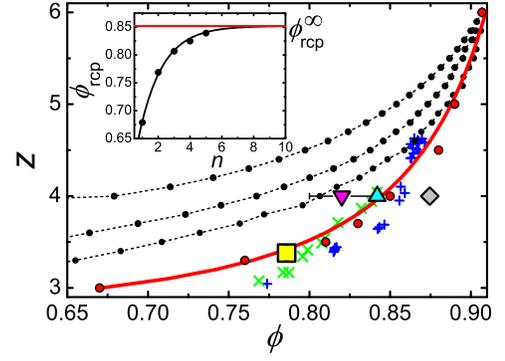}}} \caption{(Color online) {\bf Phase diagram of 2d packings.}  Theoretical results for $n=1,2,3$ (line-points, from left
to right) and $\phi^\infty$ (red) are compared to (i) values in the literature: Berryman \cite{Berryman1983} (down triangle), RT \cite{Parisi2010} (diamond), and O'Hern {\it et al}.
\cite{OHern2002} (up triangle), (ii) simulations of 10,000 monodisperse disks (crosses), and polydisperse disks (pluses) with a discrete uniform distribution of radius in $[0.7, 1.0]$
(in unit of maximum radius), and (iii) experimental data of frictional disks (square).
(inset) The theoretical RCP volume fraction $\phi_{\rm rcp}(n)$ as a function of $n$. The points are fitted to a function $\phi_{\rm rcp} (n) = \phi_{\rm rcp}^\infty - k_1 e^{-k_2n}$,
where $k_1 = 0.34 \pm 0.02$, $k_2=0.67 \pm 0.06$, and $\phi_{\rm rcp}^\infty = 0.85 \pm 0.01$ (blue dashed line). Other values of $\phi^\infty$ (with different $z$) are obtained in the
same way.} \label{fig:2Dvolume}
\end{figure}

To provide an analytical form of the single-particle angular correlation function, we consider the distribution of gaps between 1d neighboring rods. For simplicity, we only consider
the gap between rod 1 and 2 (its distribution is the same as that of other gaps due to translational invariance):
\begin{equation}
\begin{split}
G(\alpha) = &\langle \delta(x_2-x_1-\alpha)\rangle\\
=& \frac{1}{Q(L,z)}\idotsint_{0 = x_1<x_2<\cdots<x_z<L}  \prod_{i=2}^{z}dx_i \\
&\times \exp[-\beta V(x_1,\cdots, x_z)] \delta(x_2 -\alpha) \\
 =& \frac{\exp[-\beta V(\alpha)]}{Q(L,z)}\idotsint_{\alpha =
x_2<x_3<\cdots<x_z<L}\prod_{i=3}^{z}dx_i \\
&\times\exp[-\beta V(x_2,\cdots, x_z)] \\=& \frac{Q(\alpha,1)Q(L-\alpha, z-1)}{Q(L,z)}.
\end{split}
\label{eq:G}
\end{equation}

If we set $a=\infty$ in the potential $V(x)$, the system becomes a classical model -- a one dimensional gas of hard rods (Tonks gas) \cite{Tonks1936}.
In the thermodynamic limit ($L\rightarrow \infty$ and $z\rightarrow \infty$), the gap distribution is \cite{Salsburg1953, Tonks1936}:
\begin{equation}
G_{\rm HR} (\alpha) = \rho_f e^{-\rho_f (\alpha/l_0-1)}, \label{eq:GHR}
\end{equation}
where $\rho_f = z/(L/l_0-z)$ is the free density. This result is exact in 1d because the Kirkwood-like decomposition Eq.~(\ref{eq:Gn}) is satisfied. Equation~(\ref{eq:GHR}) is also
consistent with the exponential form of the distribution of Voronoi excluded volume in Ref.~\cite{Song2008}, where the 1d hard rod model is used as a mean-field approximation for 3d
packings.

\section{Phase diagram of 2d jammed packings}
\label{sec:phase_diagram}

The strategy of our method to calculate the volume fraction for a fixed coordination number $z$ is to first evaluate $G(\alpha)$ from Eq.~(\ref{eq:G}), then plug it into
Eqs.~(\ref{eq:Gn}) and~(\ref{eq:PgIso}) to calculate $P(\ell)$ and eventually obtain $\langle W \rangle_e$ and $\phi$ via Eqs. (\ref{eq:W}) and (\ref{eq:voronoi}).  Equation
~(\ref{eq:PgIso}) is a high-dimensional integration which is solved numerically by Monte Carlo method.

In the proof of the Kepler conjecture, Hales shows that considering a cluster of 50 spheres is sufficient in search for the optimal crystal packing \cite{Hales2005}. Analogously, we
expect that the volume fraction of random packings would converge quickly with $n$.
We truncate the expansion Eq.~(\ref{eq:PgIso}) to a finite value of $n$, and extrapolate the finite behavior to the infinite limit.  Indeed, our results show that $\phi(n)$ approaches
the asymptotic value $\phi^\infty$ exponentially fast as $n\to \infty$ (Fig.~\ref{fig:2Dvolume} inset).

The results can be visualized into a 2d phase diagram in the $z-\phi$ plane. Figure~\ref{fig:2Dvolume} shows the equation of state $\phi^\infty(z)$ (see Appendix~\ref{sec:theory_table}
for values)
as well as the approach to this asymptotic value for small $n$. Our formalism reproduces the highest density in 2d packings obtained by Thue and T\'{o}th \cite{Thue1892} of hexagonal
packing $\phi^\infty_{\rm hex} = 0.91$ at $z=6$.  It also predicts the densities of isostatic packings with different friction coefficients. In order to have a mechanical stable
packing, the isostatic counting argument \cite{Maxwell1864,Alexander1998} requires that $z = 2d = 4$ for frictionless packings (RCP), and $z=d+1 = 3$ for infinite frictional packings
(RLP).
Our theory asymptotically predicts in the two limiting cases: $\phi_{\rm rcp}^\infty = 0.85 \pm 0.01$ and $\phi_{\rm rlp} ^\infty = 0.67 \pm 0.01$ for $z=4$ and $z=3$, respectively.



The 2d RCP density of monodisperse packings has been estimated theoretically by Berryman from a continuous extension of the liquid phase \cite{Berryman1983}, which reports $\phi_{\rm
rcp} = 0.82 \pm 0.02$.  However, this approach is questionable due to the existence of a glass transition between liquid and jammed phases as noted in \cite{Parisi2010}. Binary disk
simulations (commonly used to suppress crystallization) obtain $\phi_{\rm rcp} \sim 0.84$ \cite{OHern2002} which is within the predicted $\phi_{\rm rcp}^\infty$. On the other hand, to
our knowledge there is no reported density of 2d RLP.

More sophisticated theories use RT to solve for the density of hard spheres \cite{Parisi2010,Krzakala2007}, and predict that packings can exist in a range of volume fractions at the
isostatic coordination number $z=2d$ \cite{Parisi2010}.  In the case of two-dimensional packings RT predicts isostatic packings in a range from the threshold density $\phi_{\rm th}=
0.8165$ to the maximum density of glass close packing $\phi_{\rm GCP} = 0.8745$~\cite{Parisi2010}.
It is interesting to interpret our prediction of a single RCP point within the range predicted by RT~\cite{Parisi2010}. The ensembles in our theory are characterized by the correlation
functions like $G_n$ or $P(\ell)$. This provides a systematic way to correlate $\phi$ to characteristic packing structures. If the isostatic packings could indeed have different
correlations, which might be protocol-dependent in the experimental realizations, then our theory would also predict multiple packing fractions as in RT, based on proper
characterizations of the correlations. This venue will test possible commonalities between Edwards statistical mechanics for jamming and the mean-field RT picture for glasses; a
unification that has been sought after in the field \cite{Krzakala2007,Mari2008,Jin2010B,Mezard2011}.

\begin{figure}
\centerline{\hbox{  \includegraphics[width=3.4in]{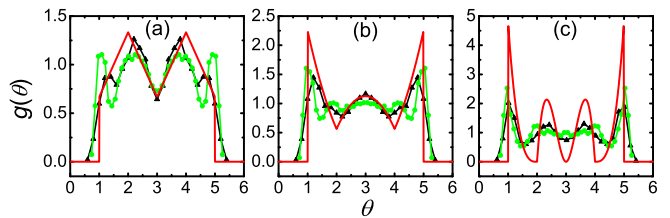}}}
 \caption{(Color online) {\bf Angular pair correlation function
  $g(\theta)$}. The theoretical $g(\theta)$ (red solid lines, rescaled by $\pi/3$) is
compared to simulation (black triangle-lines) and experimental (green circle-lines) data, with local coordination number (a) $z_1= 3$, (b) $z_1=4$, and (c) $z_1=5$~\cite{z1}.
The simulation data are obtained from a polydisperse RCP packing in order to avoid crystallization. The subset of particles with local coordination number $z_1$ is used to evaluate
$g(\theta)$.
} \label{fig:distribution}
\end{figure}

\section{Experimental and numerical tests}
\label{sec:test}

 The experiments are conducted using a granular monolayer of photoelastic disks~\cite{Puckett2013}.  The data consist of 500 packings each containing 1004 bidisperse disks in a 1:1
concentration with diameters 11.0 mm and 14.4 mm, having an interparticle friction coefficient $\mu_B \approx 0.8$. Packings are isotropically compressed and recorded using separate
images to measure the position of the disks and contact forces.
This study presents data similar to Ref.~\cite{Puckett2013}, except that we consider only the majority particles in the bath with the same friction coefficient. More experimental
details can be found in Ref.~\cite{Puckett2013} and Appendix~\ref{sec:experiments}. The average $\phi = 0.7859 \pm 0.0006$ and the average $z = 3.4 \pm 0.1$ agree well with the
prediction of the theory as seen in Fig.~\ref{fig:2Dvolume}.

Further test of the theory is obtained by comparing the correlations. For this purpose, we obtain the angular correlation function $g(\theta)$ \cite{gtheta} (equivalent to the pair
correlation function of angles, where $\theta$ is the angle between any two surface particles) from the theory (Appendix~\ref{sec:g}):
\begin{equation}
g(\theta) = \frac{L}{z} \sum_{m=1}^{z-1} \frac{Q(\theta, m) Q(l-\theta,
  z-m)}{Q(l,z)}, \label{eq:pair}
\end{equation}
which reproduces well the experimental data (Fig.~\ref{fig:distribution}). The theory deviates from data in the peak magnitudes but not locations when the local coordination number
$z_1=5$. This might be due to the presence of the polydisperse effect in the experiments (which becomes more significant for larger coordination numbers), or the neglect of higher
order correlations in the theory. The peak presented in the experimental data at $\theta \approx \pi/3$ (or $5\pi/3$) when $z_1=3$ is probably due to the remaining crystalline order in
binary packings.

We also tested the theory with simulation packings generated by the Lubachevsky-Stillinger (LS) algorithm \cite{Lubachevsky1990} and the ``split'' algorithm \cite{Song2008}. Using
simulations we are able to test the full curve of $\phi(z)$. We prepare packings for both monodisperse and polydisperse disks in the random phase $3<z<4$ by changing the interparticle
friction coefficient from zero ($z=4$, RCP) to infinity ($z=3$, RLP) \cite{simulation}. We find a good agreement except for small $z$,
which suggests that for very loose packings, higher-order correlations beyond the Kirkwood decomposition Eq.~(\ref{eq:Gn}) may be necessary. The numerical $g(\theta)$ is consistent
with theories and experiments as seen in Fig.~\ref{fig:distribution}. Our results are in line with existing analysis \cite{Troadec2002}.

\begin{figure}
\centerline{\hbox{\includegraphics[width=3.0in]{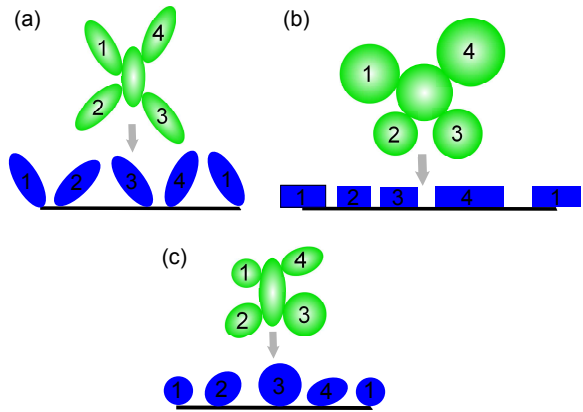}}}
\caption{(Color online) {\bf Generalization of the model.} (a) Mapping contact ellipses to the Paris car
parking model \cite{Chaikin2006}. (b) Mapping polydisperse contact disks to polydisperse rods. (c) Mapping mixtures of disks and ellipses to a 1d model.} \label{fig:mapping}
\end{figure}

\section{Discussions and conclusions}
\label{sec:conclusion}

In this paper, we construct a framework to study random packings. Our theory is based on a statistical approach, which assumes that each state can be visited with equal probability.
The approach should be applied and generalized with caution. For example, in Ref.~\cite{Puckett2013}, the authors studied the equilibrium of two subsystems with different frictions.
They found that while each subsystem is equilibrated, only the angoricity (conjugate to the stress) but not the compactivity (conjugate to the volume) equilibrates between the two
subsystems. In this case, one should appropriately integrate the stress ensemble with the volume ensemble. Moreover, in recent years, it is found that several protocols produce RCPs at
densities different from the commonly observed values ($\phi_{\rm RCP} \sim 0.64$ in 3d and $\phi_{\rm RCP} \sim 0.84$ in 2d). The ensembles generated by these protocols are likely
different from the Edwards ensemble, and the final states could depend on the dynamics of the protocols. In principle, one needs a dynamic theory for each of these protocols, and we
leave the question open whether they can be described by static theories like the present approach.

The mean field theory to Song et al. \cite{Song2008} has been generalized to particles of non-spherical shapes by Baule et al.~\cite{Baule2013, Portal2013, Baule2014}. The present
theory offers the possibility to take into account the correlations neglected in \cite{Baule2013} to build up a theory of non-spherical particles from the bottom up. For instance, 2d
packings of ellipses require a 1d model with orientations (Fig.~\ref{fig:mapping}a). The solution of such a model (named the ``Paris car parking'' problem \cite{Chaikin2006}) will lead
to a prediction of RCP and the optimal packing of
elongated particles, an open theoretical problem with implication for self-assembly of nanoparticles and liquid crystal phases. 
It is also possible to generalize this model to polydisperse systems, by explicitly calculating the dependence of the local coordination numbers with the concentration of species
\cite{Danisch2010}, and mapping the problem to a ``car parking'' problem of polydisperse cars (Fig.~\ref{fig:mapping}b). Note that in our experiments and simulations, we have
introduced a weak polydispersity to avoid crystalline order. Although one usually neglects the packing fraction corrections of weak polydispersities \cite{Troadec2002},  we expect
monodisperse theories to become insufficient for systems with strong polydispersities. Furthermore, the theory can be applied to mixtures of spherical and non-spherical objects in
search of new phases of jammed matter (Fig.~\ref{fig:mapping}c).

Overall, the present formalism facilitates a systematic investigation of correlations in packings,
and paves the path to a solvable model.
The framework may be extended to predict the optimal ordered and disordered packings over a set of specified shapes, dimensions and friction properties.

{\bf Acknowledgment.}  This work is supported by the National Science Foundation, CMMT Program and the Department of Energy, Office of Basic Energy Sciences, Geosciences Division.
 We
are grateful to C.~Song, F.~Zamponi, P.~Charbonneau, F.~Santib\'{a}\~{n}ez, K.~E.~Daniels, and R.~P.~Behringer for many useful discussions.

\appendix

\section{Calculation of angles in $P(\ell)$}

\label{sec:angles}

\begin{figure}[h!]
\centering \resizebox{1.7in}{!}
           {\includegraphics{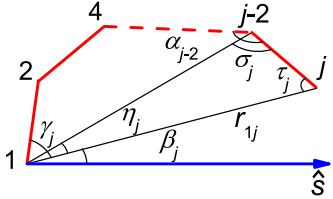}} \caption{An illustration of
             the geometrical relations between angles and
             distances.} \label{fig:angles}
\end{figure}

We need $n+1$ angles, ($\beta, \alpha_{1}, \ldots, \alpha_{n}$), to determine the positions of Voronoi particles. According to the geometrical relationships (Fig.~\ref{fig:angles}),
other angles and distances can be calculated from these integration variables recursively as:
\begin{equation}
\begin{split}
\sigma_j &= \alpha_{j-2} - \tau_{j-2} \\
r_{1j} & = \sqrt{r_{1,j-2}^2 + 1 -
  2r_{1,j-2} \cos \sigma_j} \\
\tau_j & = \arcsin
\left(\frac{r_{1,j-2}}{r_{1j}}\sin \sigma_j\right) \\
\eta_j & = \arcsin \left(\frac{1}{r_{1j}}\sin \sigma_j\right)
\\ \gamma_j & = \gamma_{j-2} + \eta_j,
\end{split}
\end{equation}
and
\begin{equation}
\beta_j = \begin{cases} \beta_2 - \gamma_j, & \mbox{if } j = 4, 6, 8
  \ldots \\ \beta_3 - \gamma_j ,& \mbox{if } j = 5, 7, 9 \ldots
\end{cases}
\end{equation}
with initial values
\begin{equation}
\begin{split}
\sigma_2 & = \tau_2 = \eta_2 = \gamma_2 = 0, \\
\beta_2 & = \beta,\\
r_{12} & = 1, \\
\sigma_3 & = \tau_3 = \eta_3 = \gamma_3 = 0, \\
\beta_3 & = \alpha_1 - \beta_2, \\
r_{13} & = 1.
\end{split}
\end{equation}

\section{A discussion on the Kirkwood-like decomposition of the $n$-particle angular correlation function}
\label{sec:kirkwood}
\begin{figure}[th!]
\centering \resizebox{1.6in}{!}
           {\includegraphics{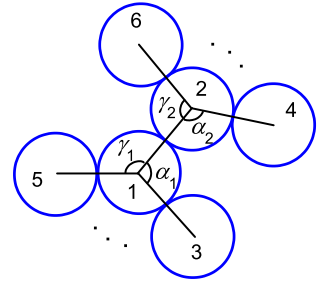}} \caption{An illustration of particles and
             angels in Eq.~(\ref{eq:G2full}). There are $z-3$ particles (not shown) between particles 3 and 5 (4 and 6).}
             \label{fig:angle}
\end{figure}
The Kirkwood-like decomposition Eq. (5) in the main text is an approximation of the $n$-particle angular correlation function $G_n(\alpha_1, \ldots \alpha_n)$, which neglects
higher-order correlations between particles that do not share a common contact neighbor. To see this, let us look at the simplest case when $n=2$. An expression of $G_2(\alpha_1,
\alpha_2)$ is:
\begin{equation}
\begin{split}
G_2(\alpha_1, \alpha_2) \sim &\int_{\gamma_1=0}^{L}\int_{\gamma_2=0}^{L} Q(\alpha_1,1) Q(\gamma_1,1)\\
&\times Q(\alpha_2,1)Q(\gamma_2,1)\\
&\times Q(L-\alpha_1-\gamma_1, z-2)\\
&\times Q(L-\alpha_2-\gamma_2, z-2)\\
&\times \Theta(r_{34}-1) \Theta(r_{56}-1) d\gamma_1 d\gamma_2,
\end{split}
\label{eq:G2full}
\end{equation}
where $L = 2 \pi$, and $Q(L,z)$ is the partition function of 1d rods (see below).
The particles and angles are indicated in Fig.~\ref{fig:angle}.
The Heaviside step functions impose the hard-sphere constraints between particles 3 and 4, and between 5 and 6, which are not in direct contact with any common neighbors (compared to
``direct" particles such as particles 2 and 3, which share a common neighbor particle 1).
If we neglect the hard-sphere constraints between these indirect particles, and only include correlations between direct particles, Eq.~(\ref{eq:G2full}) becomes
\begin{equation}
\begin{split}
G_2(\alpha_1, \alpha_2) \sim &\int_{\gamma_1=0}^{L}\int_{\gamma_2=0}^{L} Q(\alpha_1,1) Q(\gamma_1,1)\\
&\times Q(\alpha_2,1)Q(\gamma_2,1)\\
&\times Q(L-\alpha_1-\gamma_1, z-2)\\
&\times Q(L-\alpha_2-\gamma_2, z-2) d\gamma_1 d\gamma_2
\end{split}
\label{eq:G2approx}
\end{equation}
Because
\begin{equation}
\int_{\gamma_1=0}^{L} \frac{Q(L-\alpha_1-\gamma_1, z-2)Q(\gamma_1,1)}{Q(L-\alpha_1,z-1)}d\gamma_1 = 1,
\end{equation}
(same for $\gamma_2$), Eq.~(\ref{eq:G2approx}) can be further written as
\begin{equation}
\begin{split}
G_2(\alpha_1, \alpha_2) \sim & Q(\alpha_1,1)  Q(L-\alpha_1,z-1)\\
&\times Q(\alpha_2,1) Q(L-\alpha_2,z-1)\\
\sim & G(\alpha_1) G(\alpha_2).
\end{split}
\end{equation}
The above derivation shows that the 2-particle angular correlation function $G_2(\alpha_1, \alpha_2)$ can be approximated as a product of single-particle angular correlation functions,
if higher-order correlations are neglected. The same analysis can be extended to $G_n(\alpha_1, \ldots \alpha_n)$ when $n>2$.

\section{Partition function of 1d rods} \label{sec:part}
To simplify the notation, here we set the size of rods to be the unit, $l_0 = 1$. The full expressions (in the main text) are recovered by adding a proper scaling factor $1/l_0$ to the
distance parameters, such as $x$ and $L$. If we plug the potential Eq.~(\ref{eq:potential}) in $q(s)$ (Eq.~(\ref{eq:partition2})), we have
\begin{equation}
q(s) = \int_1^{a} e^{-sx} dx = \frac{e^{-s} - e^{-as}}{s},
\end{equation}
and the partition function becomes
\begin{equation}
\begin{split}
Q(L,z) = &\frac{1}{2\pi i} \int_{\gamma - i \infty}^{\gamma + i
  \infty} e^{sL} \left( \frac{e^{-s} - e^{-as}}{s} \right)^z ds \\ = &
\sum_{k=0}^{z}(-1)^k \binom{z}{k} \left\{ \frac{1}{2\pi i} \int_{\gamma - i \infty}^{\gamma + i
  \infty}\frac{e^{s[L-z-k(a-1)]}}{s^z} \right\}\\
  = &\sum_{k=0}^{\lfloor \frac{L-z}{a-1}\rfloor}(-1)^k
\binom{z}{k}\frac{[L-z-k(a-1)]^{z-1}}{(z-1)!}  \\
&\times \Theta(L-z)\Theta(az-L). \label{eq:SI_parti}
\end{split}
\end{equation}
where we have used the binomial expansion of $\left( \frac{e^{-s} - e^{-as}}{s} \right)^z$.

\section{Theoretical values of $\phi^\infty(z)$}
\label{sec:theory_table} In Table~\ref{tab:phi}, we list the extrapolated values of $\phi^\infty(z)$ evaluated from our theory (see Fig. 2).

\begin{table}[h]
\begin{ruledtabular}
\begin{tabular}{c c c c c c c c c }
  $z$ &  3.0 & 3.3& 3.5& 3.7 & 4.0& 4.5& 5.0 & 6.0\\\hline
  $\phi^\infty$ & 0.67 & 0.76 & 0.81 & 0.83 & 0.85 & 0.88 & 0.89 & 0.91 \\
\end{tabular}
\end{ruledtabular}
\caption{Theoretical values of $\phi^\infty(z)$.} \label{tab:phi}
\end{table}

\section{Collection of Experimental Data}
\label{sec:experiments} The experiments involve an assembly of 1004 bi-disperse, photoelastic disks having a diameter of 11.0 mm and 15.4 mm in equal concentration by number. Particles
are composed of photoelastic material (Vishay PhotoStress PSM-4) and are birefringent under strain so that contact forces can be calculated.  The granular monolayer rests on a nearly
frictionless surface of an air table and is confined by two immovable walls and two pistons.  The system is initially dilute and unjammed.  Two pistons bi-axially compress the system
through a series of small quasi-static steps with a size corresponding to $\Delta \Phi = 0.0009$.  At each step, separate images are recorded to measure the displacement and contact
forces. We use only data collected from jammed configurations over the range of $0.7836 < \phi < 0.7884$.   After the system has reached the maximum desired $\phi$, the pistons dilate
and the system is mixed.  This cycle is repeated ensuring generation of independent configurations. In this way, over 500 packings are obtained and analyzed. More details of the
experimental apparatus and procedures are reported in a recent paper, Ref.~\cite{Puckett2013}.

\section{Angular pair correlation function $g(\theta)$} \label{sec:g} The 2d angular pair correlation function $g(\theta)$ is equivalent to the pair correlation function in the 1d
model, which is the probability of finding a rod at a given distance $\theta$ from another rod. $g(\theta)$ is different from $G(\alpha)$ because other rods are allowed to be inside
$\theta$. Due to the translational invariance, we can choose any rod (rod 1 in this case) as the reference point:
\begin{equation}
\begin{split}
\rho g(\theta) =& \langle \sum_{k = 2}^{z} \delta(x_k-x_1-\theta) \rangle
\\
=&\frac{1}{Q(L,z)} \sum_{k=2}^{z} \idotsint_{0 =
  x_1<x_2<\cdots<x_z<L} \prod_{i=2}^{z}dx_i  \\
 &\times \exp[-\beta V(x_1,\cdots, x_z)] \delta(x_k -\theta)  \\
  =&\frac{1}{Q(L,z)} \sum_{k=2}^{z} \idotsint_{0
  = x_1<x_2<\cdots<x_k=\theta} \prod_{i=2}^{k-1}dx_i  \\
 &\times \exp[-\beta V(x_1,\cdots, x_{k-1})] \\
& \times \idotsint_{\theta =
  x_k<x_{k+1}<\cdots<x_z<L} \prod_{i=k+1}^{z}dx_i \\
&\times  \exp[-\beta V(x_k,\cdots, x_{z})]  \\
=& \sum_{k=2}^{z} \frac{Q(\theta, k-1) Q(L-\theta,
  z-k+1)}{Q(L,z)},
\end{split}
\end{equation}
where the number density $\rho = z/L$. From the last expression, the angular pair correlation function $g(\theta)$ can be written as
\begin{equation}
\begin{split}
g(\theta) &= \frac{1}{\rho} \sum_{m=1}^{z-1} g_m(\theta), \\
g_m(\theta) &= \frac{Q(\theta, m) Q(L-\theta, z-m)}{Q(L,z)}.
\end{split}
\label{equation:pair}
\end{equation}
The function $g_m(\theta)$ is the probability density of finding two contact particles at a relative angle $\theta$, such that there is exactly $m-1$ contact particles between them.
Equation~(\ref{equation:pair}) is used to calculate the theoretical $g(\theta)$ in Fig.~3.

The normalization of $g(\theta)$ is conventional:
\begin{equation}
\int_0^{L} \rho g(\theta) d\theta = z-1.
\end{equation}


\begin{thebibliography}{27}
\bibitem {z1} Note that in general the local coordination number $z_1$ is different from the global average coordination number $z$ (which is the average of $z_1$). However in
Eq.~(\ref{eq:PgIso}) we have used the approximation $z \approx z_1$.
\bibitem {gtheta} For the purpose of testing angular correlations, we find that the angular correlation function $g(\theta)$ is a better candidate than $G(\alpha)$. Note that
$g(\theta)$ is different from $G(\alpha)$ because $G(\alpha)$ only considers the angles between two neighboring contact particles. Therefore, $G(\alpha)$ is a monotonic descending
function while $g(\theta)$ has richer peak structures, as shown in Fig.~\ref{fig:distribution}.
\end{thebibliography}
\end{document}